\documentclass[aps,twocolumn,showpacs,floatfix,superscriptaddress]{revtex4-1}

\usepackage[dvips]{graphics}
\usepackage{epsfig}
\usepackage{times}
\usepackage{amsmath,amssymb}
\usepackage{amsthm}
\usepackage{subfigure,overpic}
\usepackage{mathrsfs}
\usepackage{multirow}

\def \beq {\begin{equation}}
\def \edq {\end{equation}}
\def \bes {\begin{subequations}}
\def \eds {\end{subequations}}
\def \beqn {\begin{equation*}}
\def \edqn {\end{equation*}}

\def \dag {\dagger}

\def \up {\uparrow}
\def \down {\downarrow}

\def \calh {{\cal{H}}}

\def \ket {\rangle}
\def \bra {\langle}

\begin{document}
\title{Cross thermoelectric coupling in normal-superconductor quantum dots}
\author{Sun-Yong Hwang}
\affiliation{Institut de F\'{\i}sica Interdisciplin\`aria i Sistemes Complexos
IFISC (CSIC-UIB), E-07122 Palma de Mallorca, Spain}
\author{Rosa L\'opez}
\affiliation{Institut de F\'{\i}sica Interdisciplin\`aria i Sistemes Complexos
IFISC (CSIC-UIB), E-07122 Palma de Mallorca, Spain}
\affiliation{Kavli Institute for Theoretical Physics, University of California, Santa Barbara, California 93106-4030, USA}
\author{David S\' anchez}
\affiliation{Institut de F\'{\i}sica Interdisciplin\`aria i Sistemes Complexos
IFISC (CSIC-UIB), E-07122 Palma de Mallorca, Spain}
\affiliation{Kavli Institute for Theoretical Physics, University of California, Santa Barbara, California 93106-4030, USA}


\begin{abstract}
We discuss the nonlinear current of an interacting quantum dot coupled to normal and 
superconducting reservoirs with applied voltage and temperature differences. Due to the particle-hole symmetry introduced by 
the superconducting lead, the pure (subgap) thermoelectric response vanishes. However,
we show that the Andreev bound states shift as the thermal gradient increases.
As a consequence, the $I$--$V$ characteristic can be tuned with a temperature bias if the system is simultaneously  
voltage biased. This is a cross effect that occurs beyond linear response only.
Furthermore, we emphasize the role of quasiparticle tunneling processes in the generation of high thermopower sensitivities.
\end{abstract}

\maketitle
\section{Introduction}
When a superconductor (S) material is attached to a normal (N) conductor the transfer of charges across the NS interface is dominated by Andreev processes~\cite{tin96} in which electrons (holes) are retroreflected as holes (electrons) adding (destroying) Cooper pairs into the ground state. Leakage of superconducting pairing correlations into the normal conductor has a profound impact when sandwiched quantum dots (QDs) are considered. Hence, multiple Andreev reflections lead to the formation of Andreev bound states observed in N-QD-S tunnel experiments when the normal contact acts as a probe terminal \cite{gra04,Dea10,Pillet10,Schi14}. These hybrid setups are excellent testbeds to examine the interplay between Coulomb interaction and proximity effects. In particular, N-QD-S systems exhibit remarkable conductance changes in the cotunneling regime~\cite{sun01} originated from  the occurrence of zero-bias anomalies, which at much lower temperatures arise from the competition between Kondo physics~\cite{Kondo} and the Yu-Shiba-Rusinov states (usual Andreev bound states in magnetically active platforms). Indeed, the Yu-Shiba-Rusinov states are viewed as precursors of Majorana quasiparticles in hybrid systems with large spin-orbit interaction and magnetic fields~\cite{Majorana,Mourik12}. Additionally, N-QD-S structures have been proposed as suitable sources of solid-state qubits~\cite{Hof09}.

Previously cited works cope solely with the electric response of hybrid systems. In contrast, their thermoelectrical response is much less understood. In QDs coupled to normal electrodes, the thermoelectric voltage $V_\text{th}$ generated in response to a small thermal gradient $\Delta T$ is greatly amplified across each dot resonance~\cite{mah96,sta93}, leading to large values of the Seebeck coefficient $S=-V_\text{th}/\Delta T$~\cite{gol10}.  Superconductivity can significantly alter these thermoelectric properties~\cite{cla95} and provide additional information, as demonstrated with Andreev interferometers~\cite{eom98,jac10}. Importantly, in hybrid nanostructures large values of $S$ have been envisaged
by breaking the particle-hole symmetry \cite{mac13,oza14,gia14}. Here, we focus on N-QD-S setups that indeed preserve such symmetry and thereby avoid the generation of electrical currents by thermal gradients alone. Remarkably enough,
we find that the nonlinear transport regime do exhibit a large cross thermoelectric coupling signal when both voltage and temperature shifts are present. Below, we discuss the details.
 \begin{figure}
\centering
\includegraphics[width=0.47\textwidth]{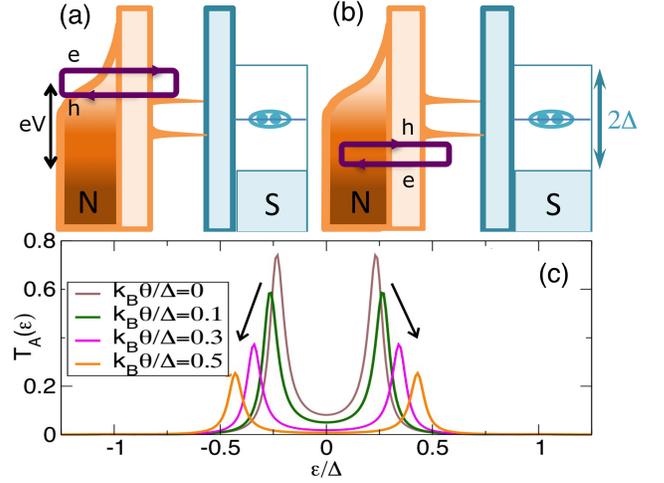}
\caption{(Color online) Top: A quantum dot device tunnel
coupled to hot normal (N) and cold superconducting (S, gap $2\Delta$) leads whose
distribution functions are sketched. Andreev processes where injected charge forms a Cooper pair in the S lead
are indicated for (a) incident electrons and (b) incident holes. For voltage bias $V=0$ both processes counterbalance
and the net Andreev current vanishes even in the presence of a thermal gradient $\theta$. 
Therefore, a cross coupling (finite $V$ and $\theta$) is needed to generate
thermoelectric transport. Bottom: Andreev transmission for different
thermal bias $\theta$. Tilted arrows indicate Andreev level shifts of the bound states indicated in (a,b) when the thermal bias enhances from $k_B\theta=0.1\Delta$ to $k_B\theta=0.5\Delta$. Parameters: $\varepsilon_d=0.15\Delta$, $\Gamma_L=0.1\Delta$, $\Gamma_R=0.5\Delta$, $k_B T=0.1\Delta$, $V=0$.}
\label{Fig1}
\end{figure}

The nonlinear regime of transport have been hitherto poorly explored in interacting hybrid systems (for an exception, see Ref.~\onlinecite{wan01}).  The nonlinear thermoelectric transport is unique because the conductor nonequilibrium potential responds in a nontrivial way to large shifts 
in the applied thermal ($\theta$) and voltage ($V$) differences~\cite{san13,whi13,mea13,lop13,her13,fah13,hwa13,dut13,whi14}.
Consider the charge current $I$ for a generic conductor,
\begin{equation}\label{eq:I2}
I=G_{0}V+G_{1}V^2+L_0 \theta+L_1\theta^2+ M_{1}V\theta\,,
\end{equation}
expanded up to leading order in the rectification terms. 
For the present setup [see top panel in Fig.~\ref{Fig1}],  all pure thermoelectrical coefficients
[$L_0$ and $L_1$ in Eq.~\eqref{eq:I2}] vanish in the subgap region due to symmetry between the processes indicated
in Fig.~\ref{Fig1}(a) and Fig.~\ref{Fig1}(b), canceling the net current. Importantly, whereas $I$ is not sensitive to a thermal gradient alone, it does react to a simultaneous change in voltage and temperature shifts because for finite $V$ and $\theta$ the process sketched in Fig.~\ref{Fig1}(b) is Pauli blocked for positive voltages. This unique cross effect is quantified to leading order by the coefficient $M_1=\partial^2 I/\partial V\partial \theta|_{\rm eq}$ (${\rm eq}$ means $\{V=\theta=0\}$). This mixed term has thus far received little attention, because its effect is masked in many systems by the diagonal responses $G_1$ and $L_1$.  We propose that hybrid systems are highly appropriate experimental setups to test this cross response as they exhibit a clear signal of $M_1$. Our results show that the $I$--$V$ characteristic can be thermally tuned due to $M_1$. Furthermore, the cross response is intimately linked with the fact that Andreev bound states shift their positions [Fig. \ref{Fig1}(c)] under the influence of a temperature shift $\theta$, a feature easily accessible with tunnel spectroscopic techniques. 

The paper is organized as follows. In Sec.~\ref{theory}, we explain the gauge-invariant nonlinear thermoelectric transport theory to describe our N-QD-S junctions. Section~\ref{results} is devoted to displaying our main results based on this theory. Importantly, the cross thermoelectric coupling term $M_1$ in Eq.~\eqref{eq:I2} plays a significant role to the Andreev current $I_A$ as illustrated in Fig.~\ref{Fig4}, where $I_A$ can increase or decrease its amplitude according to the quantum dot energy level. Furthermore, we show that high thermopower can be created when combined with the quasiparticle contribution uniquely in the nonlinear transport regime.
Finally, we conclude our findings in Sec.~\ref{conclusion}.

\section{Hybrid conductors: Nonlinear thermoelectric transport theory}\label{theory}
We consider a N-QD-S system with a heated normal lead as in Fig.~\ref{Fig1}(a). The model Hamiltonian can be written as
\begin{equation}\label{eq:ham}
 \calh=\calh_{L}+\calh_{R}+\calh_{D}+\calh_{T}\,,
 \end{equation}
 where
$\calh_{L}=\sum_{k\sigma}\varepsilon_{Lk}c_{Lk\sigma}^{\dag}c_{Lk\sigma}$ describes the electron with momentum $k$ and spin $\sigma$ in the left ($L$) normal lead while
$\calh_{R}=\sum_{p\sigma}\varepsilon_{Rp}c_{Rp\sigma}^{\dag}c_{Rp\sigma}+\sum_{p}\big[\Delta c_{R,-p\up}^{\dag}c_{Rp\down}^{\dag}+H.c.\big]$ describes electrons with momentum $p$ in the right ($R$) superconducting contact.
The second term in $\calh_R$ describes the Cooper-pair formation with energy cost $\Delta$. The dot Hamiltonian is $\calh_{D}=\sum_{\sigma}(\varepsilon_{d}-eU)d_{\sigma}^{\dag}d_{\sigma}$,
with $\varepsilon_{d}$ the dot energy level renormalized by the internal potential $U$ that accounts for electron-electron repulsion.
This interaction term will be considered at the mean-field level, which is a good approximation for metallic dots with
good screening properties~\cite{bro05}. In Eq.~\eqref{eq:ham},
$\calh_{T}=\sum_{k\sigma}t_{L}c_{Lk\sigma}^{\dag}d_{\sigma}+\sum_{p\sigma}t_{R}e^{\frac{i}{\hbar}eV_{R}t}c_{Rp\sigma}^{\dag}d_{\sigma}+H.c.$ depicts the tunneling processes between the dot and each lead
with amplitudes $t_L$ and $t_R$ ($V_R$ is the voltage in lead S).

We use a gauge-invariant current-conserving theory applied to nonlinear thermoelectric transport.  First,
the electric current can be evaluated from the time evolution of total electron number in the left lead,  $N_{L}=\sum_{k\sigma}c_{Lk\sigma}^{\dag}c_{Lk\sigma}$, through
\begin{equation}
I=-e\bra\dot{N}_{L}(t)\ket=-(ie/\hbar)\bra[\calh,N_{L}]\ket\,.
\end{equation}
Within the nonequilibrium Keldysh-Green function formalism~\cite{Cue96,Sun99}, one finds that $I=I_A+I_Q$ is a sum of two terms: the Andreev current $I_A$ and the quasiparticle contribution $I_Q$,
\begin{align}
&I_{A}=\frac{2e}{h}\int d\varepsilon~T_A(\varepsilon)\big[f_{L}(\varepsilon-eV)-f_{L}(\varepsilon+eV)\big]\label{I_A},\\
&I_{Q}=\frac{2e}{h}\int d\varepsilon~T_Q(\varepsilon)\big[f_{L}(\varepsilon-eV)-f_{R}(\varepsilon)\big].\label{I_Q}
\end{align}
Here, the Fermi-Dirac distribution function is given by $f_{\alpha=L,R}(\varepsilon\pm eV)=\{1+\exp[(\varepsilon\pm eV-E_F)/k_{B}T_{\alpha}]\}^{-1}$, where the electrode temperature $T_\alpha=T+\theta_\alpha$
is given in terms of the base temperature $T$ and the shift $\theta_\alpha$, the voltage bias reads $V=V_L-V_R$
and the Fermi level is taken as the reference energy ($E_F=0$). The Fermi function $f_R$ for the superconductor is that of the zero gap state, with the gap property included in the S density of states. Quite generally, the transmissions $T_A$ and $T_Q$ are functions of $U$, which depends itself on the applied voltage and thermal bias.

 Both $T_A$ and $T_Q$ are expressed in terms of the dot retarded Green's functions
$G^r_{ij}(\varepsilon)$ ($i,j=1,2$) in the Nambu space~\cite{Cue96}
\begin{eqnarray}
{\bf G}^r_{d}(\varepsilon)=\left(
\begin{array}{cc}
G^r_{11}(\varepsilon) & G^r_{12}(\varepsilon)\\
G^r_{21}(\varepsilon) & G^r_{22}(\varepsilon)
\end{array}
\right)\,,
\end{eqnarray}
whose matrix elements are Fourier transforms of the electronic part $G^r_{11}(t,t')=-i\Theta(t-t')\bra\{d_{\up}(t),d_{\up}^{\dag}(t')\}\ket$, the hole part $G^r_{22}(t,t')=-i\Theta(t-t')\bra\{d_{\down}^{\dag}(t),d_{\down}(t')\}\ket$, and those parts that connect both electron and hole dynamics $G^r_{12}(t,t')=-i\Theta(t-t')\bra\{d_{\up}(t),d_{\down}(t')\}\ket$, $G^r_{21}(t,t')=-i\Theta(t-t')\bra\{d_{\down}^{\dag}(t),d_{\up}^{\dag}(t')\}\ket$.
 Here, the subindex $1$ refers to the electron sector and $2$ to the hole part.
	The Green's functions relevant to our work explicitly read $G_{11}^{r}(\varepsilon)=[\varepsilon-\varepsilon_{d}+eu_L V+ez_L\theta+\frac{i\Gamma_{L}}{2}+\frac{i\Gamma_{R}}{2}\frac{|\varepsilon|}{\sqrt{\varepsilon^{2}-\Delta^{2}}}+\frac{\Gamma_{R}^{2}\Delta^{2}}{4(\varepsilon^{2}-\Delta^{2})}A^{r}(\varepsilon)]^{-1}$ and $G_{12}^{r}(\varepsilon)=G_{11}^{r}(\varepsilon)\frac{i\Gamma_{R}\Delta}{2\sqrt{\varepsilon^{2}-\Delta^{2}}}A^{r}(\varepsilon)$,
	where $A^{r}(\varepsilon)=[\varepsilon+\varepsilon_{d}-eu_L V-ez_L\theta+\frac{i\Gamma_{L}}{2}+\frac{i\Gamma_{R}}{2}\frac{|\varepsilon|}{\sqrt{\varepsilon^{2}-\Delta^{2}}}]^{-1}$
	and $u_L$ and $z_L$ will be specified below. For the subgap region $|\varepsilon|<\Delta$, one should make the substitution $\sqrt{\varepsilon^{2}-\Delta^{2}}\to i\sqrt{\Delta^{2}-\varepsilon^{2}}$.
The contribution of Andreev transmission and that of quasiparticle tunneling are respectively given by \cite{Sun99}
\begin{align}\label{eq:TA}
&T_A(\varepsilon)=\Gamma_{L}^{2}|G_{12}^{r}(\varepsilon)|^{2}\,,\\
&T_Q(\varepsilon)=\Gamma_L\widetilde{\Gamma}_R\big(|G_{11}^{r}|^{2}+|G_{12}^{r}|^{2}-\frac{2\Delta}{|\varepsilon|}\text{Re}\big[G_{11}^{r}(G_{12}^{r})^{*}\big]\big)\,,\label{eq:TQ}
\end{align}
with $\Gamma_L=2\pi|t_L|^2 \sum_k \delta (\varepsilon-\varepsilon_{Lk})$, $\Gamma_R=2\pi|t_R|^2\sum_{p}\delta(\varepsilon-\varepsilon_{Rp})$, and $\widetilde{\Gamma}_R=\Gamma_{R}\Theta(|\varepsilon|-\Delta)|\varepsilon|/\sqrt{\varepsilon^{2}-\Delta^{2}}$.
Note that $T_{A}$ describes the process where an electron (hole) incoming from the left lead is reflected as a hole (electron) backward into the same lead by producing (destroying) a Cooper pair in the S contact. Strictly at zero temperature and in the subgap regime $|eV|<\Delta$, the Andreev current is the only contribution to the total current. However, quasiparticle poisoning  ($I_Q$) must be also taken into account when temperature or voltage are sufficiently large.
The quasiparticle transmission $T_Q$ comprises the conventional tunnel processes and those in which electron or holes in the normal part become quasiparticle excitations in the superconducting reservoir, either by keeping the Cooper pair number invariant or by creating/destroying pairs~\cite{Sun99,Cue96}.

Substituting Eqs.~\eqref{eq:TA} and~\eqref{eq:TQ} in Eqs.~\eqref{I_A} and~\eqref{I_Q}, we can express the total current $I$
in terms of $G^r_{ij}(\varepsilon)$. Due to the presence of interactions, the Green function depends explicitly on the nonequilibrium screening potential $U$ which differs for each thermoelectric configuration $\{V_\alpha,\theta_\alpha\}$~\cite{san13}. Now we discuss how to determine $U$ in a system containing pairing correlations.
For a weakly nonequilibrium state, the potential can be expanded
$\delta U=\sum_{\alpha} \big[u_{\alpha} V_{\alpha}+z_\alpha \theta_{\alpha}\big]$
up to first order in the shifts $V_{\alpha}$ and $\theta_{\alpha}$. Here,
$\delta U=U-U_\text{eq}$ measures deviations of the internal potential from equilibrium and $u_\alpha=(\partial U/\partial V_\alpha)_\text{eq}$ and $z_\alpha=(\partial U/\partial \theta_\alpha)_\text{eq}$ are the characteristic potentials
describing the system response to the shifts.
Without loss of generality, we henceforth consider $V_{L}=V$, $T_{L}=T+\theta$ ($V_{R}=0$, $T_{R}=T$). 
The nonequilibrium potential satisfies the capacitance equation~\cite{note}
$\delta\rho=C\delta U$, where $C$ is the dot capacitive coupling and
the excess charge density is given by the dot distribution (lesser) Green function $\delta\rho=\rho-\rho_\text{eq}=i\int d\varepsilon [G^{<}_{11}(\varepsilon)-G^{<}_{11,\text{eq}}(\varepsilon)]$:
\begin{align}
&G_{11}^{<}(\varepsilon)=\frac{i\Gamma_{L}}{2\pi}\Big[|G_{11}^{r}|^{2}f_{L}(\varepsilon-eV)+|G_{12}^{r}|^{2}f_{L}(\varepsilon+eV)\Big]\nonumber\\
&+\frac{i\widetilde{\Gamma}_R}{2\pi}f_{R}(\varepsilon)\big(|G_{11}^{r}|^{2}+|G_{12}^{r}|^{2}-\frac{2\Delta}{|\varepsilon|}\text{Re}[G_{11}^{r}(G_{12}^{r})^{*}]\big)\,.\label{lesser11}
\end{align}
Notice that $G_{11}^{<}(\varepsilon)$ is dominated by Andreev events when energy lies within the subgap region  [identified by the term $f_{L}(\varepsilon+eV) |G_{12}^r(\varepsilon)|^{2}$ in Eq. \eqref{lesser11}]  whereas for $|\varepsilon|>\Delta$ it is dominated by \textit{(i)} the normal electronic dot distribution function [$|G_{11}^r(\varepsilon)|^{2}$ weighted by left-to-right averaged \emph{nonequilibrium} distribution function $(\Gamma_{L}f_{L}+\widetilde{\Gamma}_{R} f_{R}$)] and \textit{(ii)} the quasiparticle contribution [last terms in Eq. \eqref{lesser11}]. Importantly, the density is expressed as the sum of injected and screening charges, $\delta\rho=\rho_\text{inj}+\rho_\text{scr}$.
More explicitly, to leading order in $V$ and $\theta$ one has
$\delta\rho=\sum_{\alpha} (D_{\alpha}V_{\alpha}+\widetilde{D}_{\alpha}\theta_\alpha)-\Pi\delta U$
where the first term in brackets relates to the charge injected from lead $\alpha$ with voltage $V_\alpha$ and thermal driving $\theta_\alpha$ and the last term denotes screening effects in terms of the generalized Lindhard function $\Pi$.
We can thus define the charge injectivity $D_{\alpha}=(\partial\rho/\partial V_\alpha)_\text{eq}$, the entropic injectivity $\widetilde{D}_{\alpha}=(\partial\rho/\partial\theta_\alpha)_\text{eq}$, and the Lindhard function $\Pi=-(\delta\rho/\delta U)_\text{eq}$. These quantities, in general, contain particle and hole portions (explicit formulae are provided in Appendix~\ref{appen:A}) for which screening is calculated in the presence of superconductivity. 

Solving the capacitance equation, we find $\delta U=\sum_{\alpha} (D_{\alpha}V_{\alpha}+\widetilde{D}_{\alpha}\theta_\alpha)/(C+\Pi)$, hence we immediately have analytic expressions for the characteristic potentials, $u_\alpha=D_\alpha/(C+\Pi)$ and $z_\alpha=\widetilde{D}_\alpha/(C+\Pi)$,
in terms of the dot Green function:
\begin{align}
&u_{L}=\frac{-e\Gamma_{L}}{C+\Pi}\int\frac{d\varepsilon}{2\pi}\big(-\partial_\varepsilon f\big)
\big(|G_{11}^{r}|^{2}-|G_{12}^{r}|^{2}\big)_\text{eq}\,,\label{u_L}\\
&z_{L}=\frac{-\Gamma_{L}}{C+\Pi}\int\frac{d\varepsilon}{2\pi}\frac{\varepsilon-E_{F}}{T}\big(-\partial_\varepsilon f\big)
\big(|G_{11}^{r}|^{2}+|G_{12}^{r}|^{2}\big)_\text{eq}\,,\label{z_L}
\end{align}
where the integrands are evaluated at equilibrium. Note that we have only considered the response to the left lead since the superconductor is assumed to be at equilibrium and hence $u_RV_R+z_R\theta_R=0$ irrespective of $u_R$ and $z_R$.

\begin{figure}[t]
\centering
\includegraphics[width=0.47\textwidth,clip]{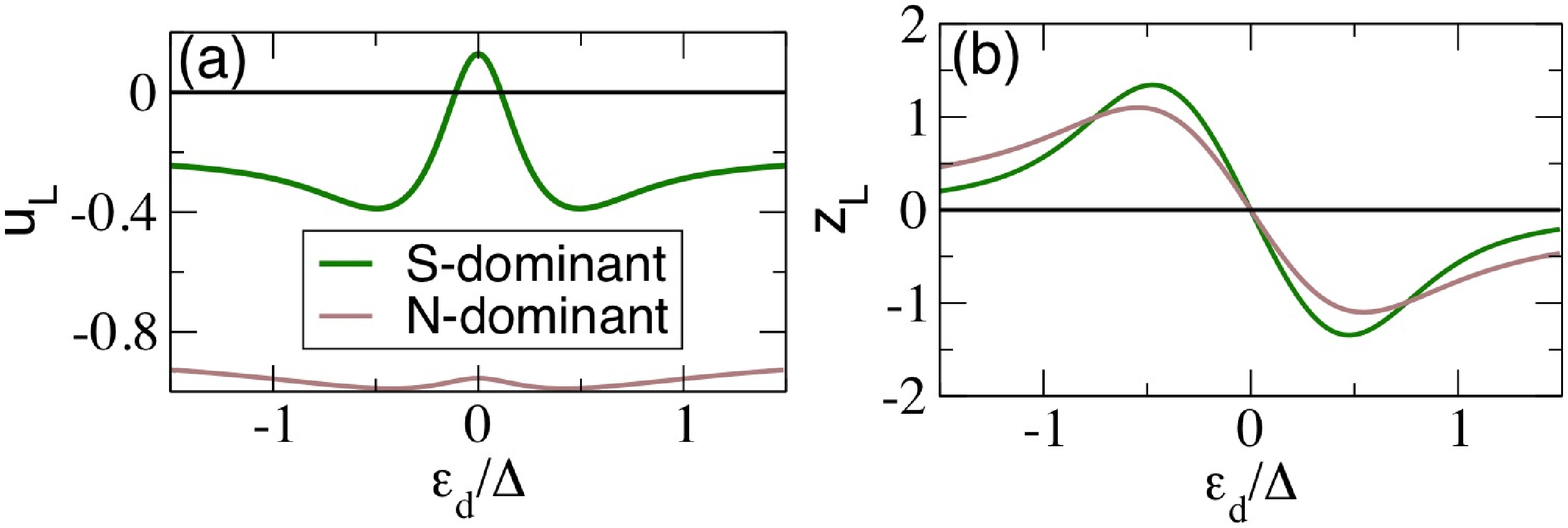}
\caption{(Color online) Characteristic potentials: (a) $u_{L}$ and (b) $z_L$ versus dot level $\varepsilon_d$ at base temperature $k_{B}T=0.1\Delta$ with $C=0$. In (a), we find $u_{L}\simeq-1$ for $\Gamma_{L}\gg\Gamma_{R}$ (N-dominant) while $u_L$ shows rather drastic change for $\Gamma_{L}\ll\Gamma_{R}$ (S-dominant). In (b), $z_L$ behaves qualitatively the same for the two cases.
}\label{Fig2}
\end{figure}

\section{Hybrid conductors: Results}\label{results}
We next discuss our main findings for the nonlinear thermoelectric transport of an interacting hybrid setup. We first focus on the internal potential changes $\delta U=u_L V+ z_L\theta$ when the hybrid system is electrically and thermally biased. These potentials are shown in Fig.~\ref{Fig2}, which displays the solutions of Eqs.~\eqref{u_L} and~\eqref{z_L} as a function of the dot level position for two coupling limits,  the normal ($\Gamma_L\gg\Gamma_R$) and superconducting ($\Gamma_L\ll\Gamma_R$) dominant cases. We consider the strongly interacting regime (in the mean-field language) by setting $C=0$. In Fig.~\ref{Fig2}(a), we find $u_{L}\simeq -1$ in the N dominant case since the applied voltage shifts the dot level as $\varepsilon_d\to\varepsilon_d-u_L eV\simeq\varepsilon_d+eV$ due to charge neutrality.
	In contrast, when the superconductor is more strongly coupled to the dot, $u_L$ behaves very differently.
	The sign changes ($u_L>0$) when particle-hole conversion process dominates over the ordinary electron tunneling, as expected from the term $(|G_{11}^{r}|^{2}-|G_{12}^{r}|^{2})_\text{eq}$ in Eq.~\eqref{u_L}.
On the other hand, when the dot is thermally driven as shown in Fig.~\ref{Fig2}(b) $z_L$ shows similar behaviors for 
both coupling limits. The effect of thermal driving $z_L$ appears only away from the particle-hole symmetry point 
when $\varepsilon_d=E_F$ (recall that $E_F=0$), as a consequence of a vanishing entropic injectivity.
Hence, $z_L$ is an odd function of $\varepsilon_d$ while $u_L$ is even.
Interestingly, the Andreev transmission $T_A$ is a function of both energy $\varepsilon$ and potential $U$ and thus
depends on voltage and $\theta$ via the characteristic potentials. This result is illustrated in Fig.~\ref{Fig1}(c)
for the case of thermal driving. Our finding thus suggests an extended controllability of 
N-QD-S junctions using thermoelectric configurations.

\begin{figure}[t]
\centering
\includegraphics[width=0.47\textwidth,clip]{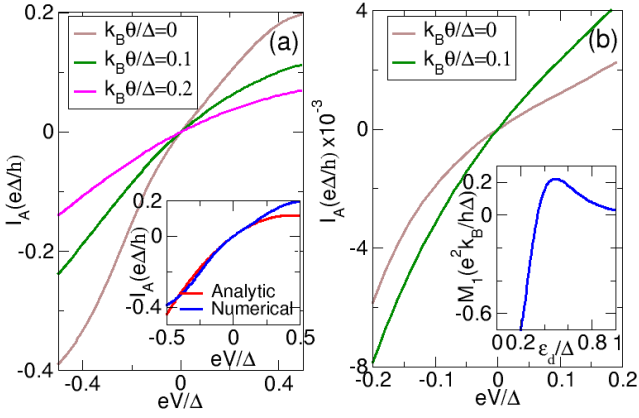}
\caption{(Color online) Andreev current versus voltage for several values of the temperature difference $k_B\theta$ with $\Gamma_L=0.1\Delta$, $\Gamma_R=0.5\Delta$ at (a) $\varepsilon_d=0.2\Delta$ and (b) $\varepsilon_d=0.7\Delta$. In (a), applied thermal bias $k_B\theta$ reduces $|I_A|$ for a given voltage since at $\varepsilon_d=0.2\Delta$, $M_1$ in Eq.~\eqref{eq:I2} is negative as shown in the inset of (b). On the contrary, at $\varepsilon_d=0.7\Delta$ as displayed in (b), $\theta$ increases the amplitude of $I_A$ due to the positive contribution from $M_1$. Inset in (a) compares Eq.~\eqref{eq:I2} with the exact expression Eq.~\eqref{I_A} at zero thermal bias. We take $k_{B}T=0.1\Delta$ and $C=0$.
}\label{Fig4}
\end{figure}

At low temperature with a small voltage bias $|eV|<\Delta$, the Andreev process largely contributes to the total current. Then, an expansion of Eq.~\eqref{I_A} as in Eq.~\eqref{eq:I2} gives the linear conductance
\begin{equation}
G_{0}=\frac{4e^{2}}{h}\int d\varepsilon \big(-\partial_\varepsilon f\big)T_{A,\text{eq}}\,,
\end{equation}
the leading-order rectification term
\begin{equation}
G_{1}=\frac{4e^{2}}{h}\int d\varepsilon \big(-\partial_\varepsilon f\big)u_{L}\frac{dT_{A}}{dU}\bigg|_{\text{eq}}\,,
\end{equation}
and the cross thermoelectric coupling
\beq
M_{1}=\frac{4e^{2}}{h}\int d\varepsilon\big(-\partial_\varepsilon f\big)\bigg[z_{L}\frac{dT_{A}}{dU}
+\frac{\varepsilon-E_{F}}{T}\frac{\partial T_{A}}{\partial\varepsilon}\bigg]_{\text{eq}}\,.\label{eq:M1}
\edq
As we anticipated above, the thermoelectric response for the subgap transport regime $(\partial I_A/\partial\theta)_{V=0}$ vanishes to any order in $\theta$, as a result of the Andreev current expression  [see Eq.~\eqref{I_A}].
Then, $L_0=L_1=0$ and Eq.~\eqref{eq:I2} simply becomes $I=G_0 V+G_1V^2+M_1V\theta$ ($I_Q\simeq 0$ for $|eV|<\Delta$). We therefore predict that the leading-order thermal response of a N-QD-S system is completely determined by the cross thermoelectric coefficient $M_1$. By turning on the electrical bias  (provided that $\varepsilon_d\neq 0$) electron-hole symmetry is lifted and then a finite response to a thermal bias is expected. 
This feature is unique to Andreev processes since normal tunneling, quite generally, gives nonzero $L_0$ and $L_1$~\cite{san13}.

Our finding is now illustrated in Fig.~\ref{Fig4}. The effect of interactions is clearly visible as $I_A-V$ is rectified for nonzero values of $\varepsilon_d\ne 0$: (a) $\varepsilon_d=0.2\Delta$ and (b) $\varepsilon_d=0.7\Delta$.
Since the expansion in Eq.~\eqref{eq:I2} is only valid within the bias range $|eV/\Delta|\ll|G_0/G_1|$ and $|k_B\theta/\Delta|\ll|G_0/M_1|$, we also compare in the inset of Fig.~\ref{Fig4}(a) with the exact expression in Eq.~\eqref{I_A}. Importantly, the $I_A$--$V$ curves in Fig.~\ref{Fig4}  (we choose the S dominant case since it offers the clearest signal) can be manipulated with the heating gradient $\theta$. The precise behavior of $I_A-V$ with $\theta$ depends on the dot gate position by either increasing [Fig.~\ref{Fig4}(b)] or decreasing [Fig.~\ref{Fig4}(a)] the amplitude of $I_A$ with increasing $\theta$. To leading order in $V$ and $\theta$, this effect is explained by the cross coupling coefficient $M_1$. A salient feature of $M_1$ is its dependence of interactions through the characteristic potential $z_L$ in competition with the noninteracting contribution given by the second term in the integrand of Eq.~\eqref{eq:M1}. The inset of Fig.~\ref{Fig4}(b) shows $M_1$ as a function of $\varepsilon_d$ in the S dominant case, showing that the sign of $M_1$ can be tuned by a dot gate potential. This sign determines the lowering or raising of $I_A-V$ with $\theta$. We also note that although the thermal bias highly affects $I_A$, the current never vanishes except for the trivial point $V=0$. This differs from the N-QD-N case~\cite{sta93,fah13}. Hence, a thermovoltage in a N-QD-S system can be created via $I_Q$ only, an important result which we discuss below. 

\begin{figure}[t]
\centering
\includegraphics[width=0.47\textwidth,clip]{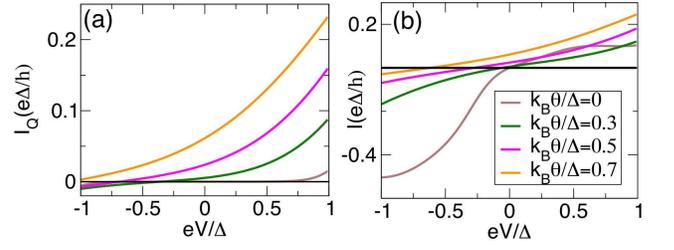}
\caption{(Color online) (a) Quasiparticle and (b) total current versus voltage for several $k_B\theta$ at $\varepsilon_d=0.3\Delta$ for $\Gamma_L=0.1\Delta$, $\Gamma_R=0.5\Delta$. Thermal agitation makes $I_Q$ highly contributing to $I$ even for a small bias range $|eV|,|k_B\theta|\ll\Delta$. For a finite  $k_B\theta\ne0$ nontrivial solutions of $I_Q=0$ and $I=0$ appear (to facilitate the localization of  $V_\text{th}$ from both $I_Q-V$ and $I-V$ curves we have drawn a horizontal black line at zero y-axis value). We take $k_BT=0.1\Delta$.
}\label{Fig6}
\end{figure}

	Unlike the Andreev current discussed above, the expansion of the quasiparticle current in Eq.~\eqref{I_Q} includes all linear and lowest order nonlinear coefficients, $I_Q=G_{0}^{Q}V+G_{1}^{Q}V^{2}+L_{0}^{Q}\theta+L_{1}^{Q}\theta^{2}+M_{1}^{Q}V\theta$.
	For low $T$, $I_Q$ is a small contribution when $|eV|<\Delta$. 
	However, as the thermal bias $\theta$ increases $I_Q$ starts to contribute nontrivially to the total current $I$ even for a small voltage bias $|eV|<\Delta$, see Fig.~\ref{Fig6}(a). Importantly, we now find 
	finite values of the thermovoltage $V_\text{th}$
determined from the condition $I_Q(V_\text{th})=0$. This is seen in Fig.~\ref{Fig6}(a) as $I_Q$ left shifts with growing $\theta$, mainly due to nonzero $L_0$ (which occurs for $\varepsilon_d\ne 0$). 
	Furthermore, the thermovoltage sign can be controlled by the gate potential polarity, i.e., positive (negative) values of $\varepsilon_d$ generates negative (positive) $V_\text{th}$. Even if the Andreev current cannot exhibit the Seebeck effect solely by itself (because no thermovoltage is generated) $I_A$ shifts the value of $V_\text{th}$ when the total current $I=I_A+I_Q$ is considered, see Fig.~\ref{Fig6}(b). Indeed, $I_A$ plays a significant role in the suppression of thermopower at low thermal biases.
	
\begin{figure}[t]
\centering
\includegraphics[width=0.47\textwidth,clip]{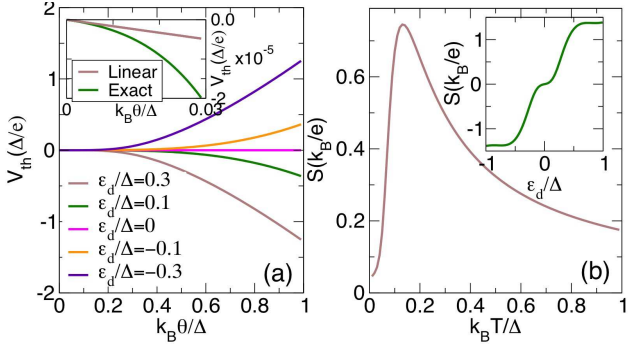}
\caption{(Color online) (a) Thermovoltage versus $k_B\theta/\Delta$ for several $\varepsilon_d$. (b) Differential Seebeck coefficient  versus base temperature at $\varepsilon_d=k_B\theta=0.3\Delta$, with $\Gamma_L=0.1\Delta$ and $\Gamma_R=0.5\Delta$. As shown in the inset of (a), the linear regime is very narrow suggesting that the observed thermoelectric effect  is intrinsically nonlinear. Inset of (b) displays the thermopower versus the dot level for $k_BT=0.1\Delta$ and $k_B\theta=0.3\Delta$.
}\label{Fig7}
\end{figure}

	Figure~\ref{Fig7}(a) shows the created thermovoltage $V_\text{th}$ from the total current $I$ for the S dominant case. We find that $V_\text{th}$ is vanishingly small until a certain amount of $\theta$ is applied, after which $V_\text{th}$ boosts. We attribute the suppression of $V_\text{th}$ around $k_B\theta<0.2\Delta$ to the Andreev current, which lacks the coefficient of $\theta^n$ to any order $n$ as discussed above.
	We check this in the inset of Fig.~\ref{Fig7}(a), which compares the linear response thermovoltage evaluated from $I=(G_0+G_0^{Q})V_\text{th}+L_0^Q\theta=0$  with the exact expression from the sum of Eqs.~\eqref{I_A} and \eqref{I_Q}.
Clearly, the linear approximation quickly fails, implying that the thermopower in our system is inherently nonlinear.
	In Fig.~\ref{Fig7}(b), we show the nonlinear Seebeck coefficient $S=-(dV/d\theta)_{I=0}$ at $\varepsilon_d=k_B\theta=0.3\Delta$
	(a moderate value of the thermal gradient). We observe a high degree of tunability for $S$
as a function of the background temperature. Another tuning parameter is the dot level. The inset of Fig.~\ref{Fig7}(b)
shows that $S$ sharply increases as $\varepsilon_d$ is detuned from the particle-hole symmetric point. Importantly, the observed thermoelectric effects can appear with all relevant energy scales well below $\Delta$.  

\section{Final remarks}\label{conclusion}
The combined influence of applied voltages and temperature biases is a fundamental aspect of electric transport.
We have here discussed interaction-driven thermoelectric effects
appearing uniquely in the nonlinear transport regime of a normal--quantum-dot--superconducting system.
Andreev processes cancel linear Seebeck effects and a nonlinear treatment of thermopower is thus called for.
We have demonstrated that a mixed thermoelectric response determines the thermal driving of Andeev currents.
This result will be robust in the presence of strong interactions (Coulomb blockade) since even when the charging
energy is fixed the cross thermoelectric coefficient will be nonzero if the transmission is energy
dependent, as in most quantum-dot setups.
In addition, we have found that high thermovoltages can be generated due to quasiparticle tunneling for moderate thermal
gradients, an effect which can have crucial importance
for superconducting-based thermometry and cooling applications~\cite{rmp}.

\begin{acknowledgments}
This research was supported by MINECO under Grant No. FIS2011-23526, the Korean NRF-2014R1A6A3A03059105, and in part by the Kavli Institute for Theoretical Physics through NSF grant PHY11-25915.
\end{acknowledgments}

\appendix

\section{Injectivities and Lindhard function}\label{appen:A}
	Using the definitions displayed in the main article, we determine the charge and entropic injectivities $D_{L}=(\partial\rho/\partial V_L)_\text{eq}$, $\widetilde{D}_{L}=(\partial\rho/\partial\theta_L)_\text{eq}$, and the Lindhard function $\Pi=-(\delta\rho/\delta U)_\text{eq}$.
	We separate the density function ($\rho=\rho^p+\rho^h$) into particle ($\rho^p$) and hole ($\rho^h$) sectors according to the Nambu space matrix elements $G_{11}^r(\varepsilon)$ and $G_{12}^r(\varepsilon)$. We find ($D_L=D_L^p+D_L^h$, $\widetilde{D}_L=\widetilde{D}_L^p+\widetilde{D}_L^h$, and $\Pi=\Pi^{p}+\Pi^{h}$)
\begin{align*}
&D_{L}^{p}=\frac{\partial\rho^{p}}{\partial V}\bigg|_{\text{eq}}
	=-e\Gamma_{L}\int\frac{d\varepsilon}{2\pi}\big(-\partial_\varepsilon f\big)
\big|G_{11,\text{eq}}^{r}(\varepsilon)\big|^{2},\\
&D_{L}^{h}=\frac{\partial\rho^{h}}{\partial V}\bigg|_{\text{eq}}
	=e\Gamma_{L}\int\frac{d\varepsilon}{2\pi}\big(-\partial_\varepsilon f\big)
\big|G_{12,\text{eq}}^{r}(\varepsilon)\big|^{2},\\
&\widetilde{D}_{L}^{p}=\frac{\partial\rho^{p}}{\partial\theta}\bigg|_{\text{eq}}
	=-\Gamma_{L}\int\frac{d\varepsilon}{2\pi}\frac{\varepsilon-E_{F}}{T}\big(-\partial_\varepsilon f\big)
\big|G_{11,\text{eq}}^{r}(\varepsilon)\big|^{2},\\
&\widetilde{D}_{L}^{h}=\frac{\partial\rho^{h}}{\partial\theta}\bigg|_{\text{eq}}
	=-\Gamma_{L}\int\frac{d\varepsilon}{2\pi}\frac{\varepsilon-E_{F}}{T}\big(-\partial_\varepsilon f\big)
\big|G_{12,\text{eq}}^{r}(\varepsilon)\big|^{2},
\end{align*}

\begin{widetext}
\begin{align*}
\Pi^{p}&=-\frac{\delta\rho^p}{\delta U}\bigg|_{\text{eq}}=\int\frac{d\varepsilon}{2\pi}f_{\text{eq}}(\varepsilon)\Bigg[
\Gamma_{L}\frac{\delta\big|G_{11}^{r}(\varepsilon)\big|^{2}}{\delta U}
+\widetilde{\Gamma}_{R}\bigg(
\frac{\delta\big|G_{11}^{r}(\varepsilon)\big|^{2}}{\delta U}
-\frac{\Delta}{|\varepsilon|}\frac{\delta}{\delta U}G_{11}^{r}\big[G_{12}^{r}\big]^{*}\bigg)\Bigg]_{\text{eq}},\\
\Pi^{h}&=-\frac{\delta\rho^h}{\delta U}\bigg|_{\text{eq}}=\int\frac{d\varepsilon}{2\pi}f_{\text{eq}}(\varepsilon)\Bigg[
\Gamma_{L}\frac{\delta\big|G_{12}^{r}(\varepsilon)\big|^{2}}{\delta U}
+\widetilde{\Gamma}_{R}\bigg(
\frac{\delta\big|G_{12}^{r}(\varepsilon)\big|^{2}}{\delta U}
-\frac{\Delta}{|\varepsilon|}\frac{\delta}{\delta U}G_{12}^{r}\big[G_{11}^{r}\big]^{*}\bigg)\Bigg]_{\text{eq}}.
\end{align*}
\end{widetext}
	For a normal conductor ($\Delta=0$) at zero temperature ($T=0$) and close to resonance ($\varepsilon_d\simeq E_F$), we obtain the Breit-Wigner-like approximation for the Lindhard function, $\Pi=(e/\pi)(\Gamma/2)[(\varepsilon_d-E_F)^{2}+(\Gamma/2)^{2}]^{-1}$ with $\Gamma=\Gamma_L+\Gamma_R$.


\begin{thebibliography}{99}
\bibitem{tin96}
M. Tinkham, \textit{Introduction to Superconductivity} (McGraw-Hill, Singapore, 1996).
\bibitem{gra04}
M. R. G\"aber, T. Nussbaumer, W. Belzig, and C. Sch\"onenberger, Nanotechnology \textbf{15}, S479 (2004).
\bibitem{Dea10}
R. S. Deacon, Y. Tanaka, A. Oiwa, R. Sakano, K. Yoshida, K. Shibata, K. Hirakawa, and S. Tarucha, Phys. Rev. Lett \textbf{104}, 076805 (2010); Phys. Rev. B  \textbf{81}, 121308(R) (2010).
\bibitem{Pillet10}
 J.-D. Pillet, C. H. L. Quay, P. Morfin, C. Bena, A. L. Yeyati, and P. Joyez, Nature Phys.\textbf{ 6}, 965 (2010).
 \bibitem{Schi14}
J. Schindele, A. Baumgartner, R. Maurand, M. Weiss, and C. Sch\"onenberger,
Phys. Rev. B \textbf{89}, 045422  (2014).
\bibitem{sun01}
Q.-F. Sun, H. Guo, and T. H. Lin, Phys. Rev. Lett. \textbf{87}, 176601 (2001).
\bibitem{Kondo}
R. Fazio and R. Raimondi, Phys. Rev. Lett. \textbf{80}, 2913 (1998);
K. Kang, Phys. Rev. B \textbf{58}, 9641 (1998);
P. Schwab and R. Raimondi, Phys. Rev. B \textbf{59}, 1637 (1999);
A. A. Clerk, V. Ambegaokar, and S. Hershfield, Phys. Rev. B \textbf{61}, 3555 (2000);
J. C. Cuevas, A. Levy Yeyati, and A. Mart\'{\i}n-Rodero, Phys. Rev. B \textbf{63}, 094515 (2001);
Y. Tanaka, N. Kawakami, and A. Oguri, J. Phys. Soc. Jpn. \textbf{76}, 074701 (2007);
Y. Avishai, A. Golub, and A. D. Zaikin, Phys. Rev. B \textbf{63}, 134515 (2001).
\bibitem{Majorana}
J. D. Sau, R. M. Lutchyn,  S. Tewari, and S. Das Sarma, Phys. Rev. Lett. \textbf{104}, 040502 (2010); R. M. Lutchyn, J. D. Sau, and S. Das Sarma,  Phys. Rev. Lett. \textbf{105}, 077001 (2010); Y. Oreg, G. Refael, and F. von Oppen, Phys. Rev. Lett. \textbf{105}, 177002 (2010).
\bibitem{Mourik12}
V. Mourik, K. Zuo, S. M. Frolov, S. R. Plissard, E. P. A. M. Bakkers, and L. P. Kouwenhoven, Science \textbf{336}, 1003 (2012).
\bibitem{Hof09}
L. Hofstetter, S. Csonka, J. Nygard, and  C. Sch\"{o}̈nenberger, Nature \textbf{461}. 960 (2009); L. G. Herrmann, 
F. Portier, P. Roche, A. Levy Yeyati, T. Kontos, and C. Strunk, Phys. Rev. Lett \textbf{104}, 026801 (2010);
A. Das, Y. Ronen, M. Heiblum, D. Mahalu, A. V. Kretinin, and H. Shtrikman, Nature Commun. \textbf{3}, 1165 (2012).
\bibitem{mah96}
G. D. Mahan and J. O. Sofo, Proc. Natl. Acad. Sci. USA \textbf{93}, 7436 (1996).
\bibitem{sta93}
A. A. M. Staring, L. W. Molenkamp, B. W. Alphenaar,
H. van Houten, O. J. A. Buyk, M. A. A. Mabesoone, C.
W. J. Beenakker, and C. T. Foxon, Europhys. Lett. \textbf{22},
57 (1993).
\bibitem{gol10}
H. J. Goldsmid, \textit{Introduction to Thermoelectricity} (Springer-Verlag, Berlin, 2010).
\bibitem{cla95}
N. R. Claughton, M. Leadbeater, and C. J. Lambert, J. Phys.: Condens. Matter \textbf{7}, 8757 (1995).
\bibitem{eom98}
J. Eom, C.-J. Chien, and V. Chandrasekhar, Phys. Rev. Lett. {\bf 81}, 437 (1998).
\bibitem{jac10}
P. Jacquod and R. S. Whitney, EPL \textbf{91}, 67009 (2010).
\bibitem{mac13}
P. Machon, M. Eschrig, and W. Belzig, Phys. Rev. Lett. {\bf 110}, 047002 (2013).
\bibitem{oza14}
A. Ozaeta, P. Virtanen, F. S. Bergeret, and T. T. Heikkil\"a, Phys. Rev. Lett. {\bf 112}, 057001 (2014).
\bibitem{gia14}
F. Giazotto, J. W. A. Robinson, J. S. Moodera, and F. S. Bergeret, Appl. Phys. Lett. \textbf{105}, 062602 (2014).

\bibitem{wan01}
J. Wang, Y. Wei, H. Guo, Q.-F. Sun, and T.-H. Lin, Phys. Rev. B \textbf{64}, 104508 (2001).

\bibitem{san13} 
D. S\'anchez and R. L\'opez, Phys. Rev. Lett. {\bf 110}, 026804 (2013).
\bibitem{whi13}
R. S. Whitney, Phys. Rev. B \textbf{87}, 115404 (2013).
\bibitem{mea13}
J. Meair and P. Jacquod, J. Phys.: Condens. Matter \textbf{25}, 082201 (2013).
\bibitem{lop13}
R. L\'opez and D. S\'anchez, Phys. Rev. B \textbf{88}, 045129 (2013).
\bibitem{her13}
S. Hershfield, K. A. Muttalib, and B. J. Nartowt,
Phys. Rev. B \textbf{88}, 085426 (2013).
\bibitem{fah13}
S. Fahlvik Svensson, E. A. Hoffmann, N. Nakpathomkun, P. M. Wu, H. Q. Xu, H. A. Nilsson, D. S\'anchez,
V. Kashcheyevs and H. Linke, New J. Phys. \textbf{15}, 105011 (2013).
\bibitem{hwa13}
S.-Y. Hwang, D. S\'anchez, M. Lee, and R. L\'opez,
New J. Phys. \textbf{15}, 105012 (2013); Phys. Rev. B \textbf{90}, 115301 (2014).
\bibitem{dut13}
P. Dutt and K. Le Hur,
Phys. Rev. B \textbf{88}, 235133 (2013).
\bibitem{whi14}
R. S. Whitney, Phys. Rev. Lett. {\bf 112}, 130601 (2014).

\bibitem{bro05}
P. W. Brouwer, A. Lamacraft, and K. Flensberg,
Phys. Rev. B \textbf{72}, 075316 (2005).

\bibitem{Cue96}
J. C. Cuevas, A. Mart\'{\i}n-Rodero, and A. Levy Yeyati, Phys. Rev. B \textbf{54}, 7366 (1996).
\bibitem{Sun99}
Q.-F. Sun, J. Wang, and T.-H. Lin, Phys. Rev. B \textbf{59}, 3831 (1999).

\bibitem{note}
S. Pilgram, H. Schomerus, A. M. Martin, and M. B\"uttiker, Phys. Rev. B \textbf{65}, 045321 (2002).
This capacitance model is a good approximation for homogeneous (zero-dimensional) quantum dots. The case
of nonuniform fields can be straightforwardly implemented in our approach but is computationally
daunting.


\bibitem{rmp}
F. Giazotto, T. T. Heikkil\"a, A. Luukanen, A. M. Savin, and J. K. Pekola,
Rev. Mod. Phys. \textbf{78}, 217 (2006).

\end{thebibliography}
\end{document}